\documentclass[aps,prl,superscriptaddress]{revtex4-1}
\usepackage{graphicx}
\usepackage{color}

\begin{document}

\title{Stretched and compressed exponentials in the relaxation dynamics
of a metallic glass-forming melt}


\author{Zhen Wei Wu}
\email[]{zwwu@pku.edu.cn}
\affiliation{International Center for Quantum Materials,
School of Physics, Peking University, Beijing 100871, China}
\affiliation{School of Systems Science, Beijing Normal University,
Beijing 100875, China}
\author{Walter Kob}
\email[]{walter.kob@umontpellier.fr}
\affiliation{Laboratoire Charles Coulomb, University of Montpellier
and CNRS, 34095 Montpellier, France}
\author{Wei-Hua Wang}
\affiliation{Institute of Physics, Chinese Academy of Sciences,
Beijing 100190, China}
\author{Limei Xu}
\email[]{limei.xu@pku.edu.cn}
\affiliation{International Center for Quantum Materials,
School of Physics, Peking University, Beijing 100871, China}
\affiliation{Collaborative Innovation Center of Quantum Matter, Beijing, China}

\date{\today}



\begin{abstract} 
The relaxation dynamics of glass-forming systems shows a multitude of
features that are absent in normal liquids, such as non-exponential
relaxation and a strong temperature-dependence of the relaxation time.
Connecting these dynamic properties to the microscopic structure of the
system is very challenging because of the presence of the structural
disorder. Here we use computer simulations of a metallic glass-former to
establish such a connection. By probing the temperature and wave-vector
dependence of the intermediate scattering function we find that the
relaxation dynamics of the glassy melt is directly related to the local
arrangement of icosahedral structures: Isolated icosahedra give rise to a
flow-like stretched exponential relaxation whereas clusters of icosahedra
lead to a compressed exponential relaxation that is reminiscent to the
one found in a solid. Our results show that in metallic glass-formers
these two types of relaxation processes can coexist and give rise to a
surprisingly complex dynamics, insight that will allow to get a deeper
understanding of the outstanding mechanical properties of metallic
glasses.

\end{abstract}


\maketitle

The relaxation dynamics of glass-forming
liquids has been and still is a topic of intensive
research~\cite{Deb01,Bin11,varshneya_06}. This activity
is motivated by the fact that glasses are not only important for a
multitude of industrial and daily applications but also pose a formidable
intellectual challenge since so far there is no theoretical framework
that is able to give a satisfactory description of the many unusual
features of glass-forming liquids and glasses. E.g.~most glass-forming
systems show in the liquid states a stretched exponential decay of
their time correlation functions~\cite{Edi96,Kob95,Kob99},
and it is believed that this is directly related to the so-called
dynamical heterogeneities~\cite{Edi96,Kob95,Kob99,Puo17} in the system,
a phenomenon for which at present there is no solid theoretical
understanding~\cite{berthier_11}. For temperatures slightly below
the glass transition temperature $T_g$ the systems show instead {\it
compressed} exponentials which are speculated to be related to the release
of internal stresses~\cite{Rut14,Rut12,Cip00,Bal08,Car08,Guo09}, although
also in this case we lack a good understanding for this behavior. Note
that this type of stress relaxation is expected to be also important deep
in the glass state~\cite{Rut14,Rut12} and thus to be related to the mechanical
properties of the material, hence giving a rational why certain glasses,
such as metallic glasses, are ductile and others, e.g.~oxide glasses,
are brittle~\cite{ciccotti_09,sun_15}.

Since in glass physics all dynamical features are a smooth function
of temperature it must be expected that such internal stresses are
in fact already present in the deeply supercooled melt and hence the
compressed exponentials should be observable to some extent already at
temperatures above $T_g$, i.e.~in the glassy melt. However, detecting
in experiments the coexistence of such different relaxation behavior
is not an easy task, since one first needs to identify observables that
are related to the different relaxation mechanisms and then be able to
measure their corresponding correlation functions. Furthermore it cannot
be expected that all glass-forming systems will show such a coexistence
and hence a wise choice on the material has to be made. Since for the
case of metallic glasses one does indeed find a cross-over from stretched
to compressed relaxation if temperature is decreased~\cite{Rut12,Luo17},
such systems seem to be good candidates to detect the simultaneous
presence of both relaxation mechanisms, and in the following we show
that this expectation is indeed borne out.

For metallic glasses it is well documented that local icosahedral clusters
are important structural building blocks~\cite{lad_12,jaiswal_15} and it has
been found that atoms in these entities have a smaller than average
atomic volume, a higher than average elastic modulus,
and show a slower than average dynamics~\cite{Wak10,Wu13a,Li09,li_17}.
These properties indicate that these clusters are indeed relevant for the
slowing down of the dynamics of the melt and the mechanical properties
of the glass~\cite{She06}. In the following we will thus focus on atoms that
are at the center of these icosahedra and in particular will consider various
types of clusters formed by such interlocked icosahedra, i.e.~structures that
have a size larger than the typical interatomic distance. By probing the
wave-vector dependence of the relaxation of the intermediate scattering
function for different clusters we find that the relaxation dynamics shows
a stretched exponential time dependence for weakly connected clusters
whereas strongly connected clusters have a compressed exponential time
dependence, thus demonstrating that for metallic glasses these two types
of processes can indeed coexist, a behavior that we regard to be responsible 
for the outstanding mechanical properties of these systems.

\begin{figure}[ht]
\centering
\includegraphics[width=0.6\linewidth]{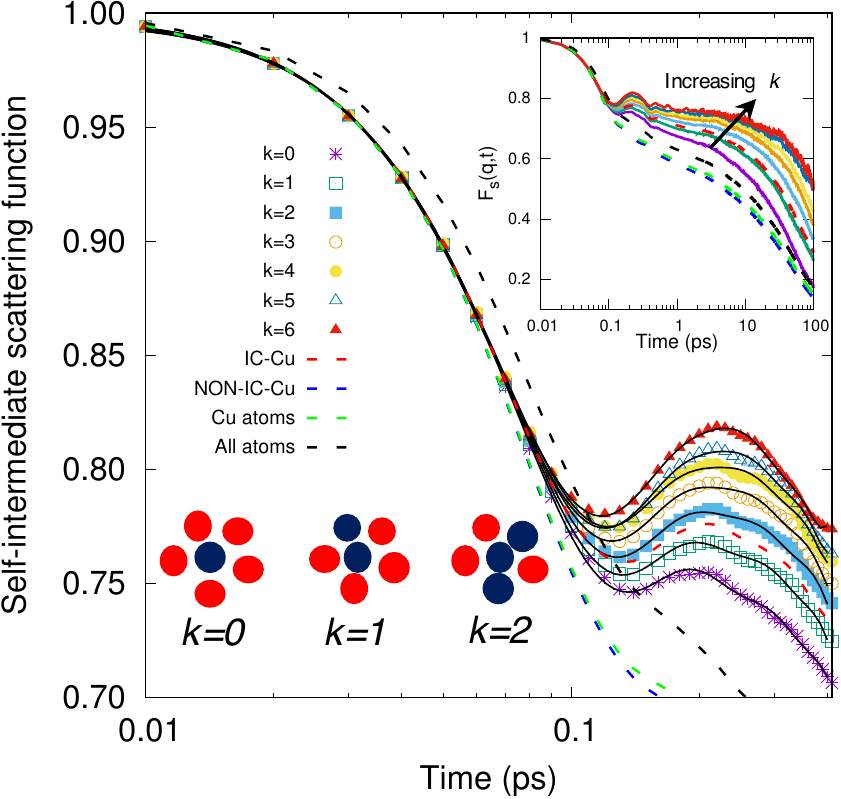}
\caption{
{\bf Relaxation dynamics of the system at short times.} Short time
behavior of the self intermediate scattering function of particles
with different local connectivity $k$ (symbols). The wave-vector is
$q=2.8$~\AA$^{-1}$ and $T=1000$K. With increasing $k$ the height of the peak at around
0.2~ps increases showing that the motion becomes less damped. The solid
lines are fits to the data with Eq.~(\ref{eq2_fsqfit}). Also included is
$F_s(q,t)$ for the average Cu atom in an icosahedral cluster (dashed red
line), the average Cu atom not in an icosahedral cluster (blue dashed
line), and the average Cu atoms (green). The black dashed line is the
correlation function average over all atoms.  The upper inset shows
the same data in a larger time interval. The lower inset illustrates
the definition of particles with different connectivities $k$: Particles in
blue are the center of an icosahedral-like cluster.}
\label{fig1_fsqt}
\end{figure}

The studied Cu$_{50}$Zr$_{50}$ system consists of $N=10000$ atoms
interacting via an embedded atom potential~\cite{Men07}. (See
Methods for details on the simulations.) For this system the
onset temperature, i.e.~the point at which the dynamics becomes
glassy~\cite{Bin11}, is around 1250~K (see SI), and in the following
we study the properties of the system between 950K and 1100K, i.e.~in
a $T-$range in which the dynamics is already rather glassy (see SI
for the $T-$dependence of the relaxation time.). As discussed above,
in this system icosahedral-like clusters with a Cu atom in its center
constitute slowly relaxing structures~\cite{Wu13a,li_17,lad_12} and
hence we focus here on the dynamics of these entities. For this we
use a Voronoi construction~\cite{Fin77,She06} to identify particles
that are 12-fold coordinated and divide them into groups defined by
the number of Cu atoms, $k$, that are in the nearest neighbor shell
of the central atoms and are also in the center of an icosahedron
(see lower inset of Fig.~\ref{fig1_fsqt})~\cite{Wu13a,Wu16}. In the
following we will refer to these Cu atoms as particles with connectivity
$k$ and in the SI we show how the concentration of these populations
change with temperature $T$. The space and time dependence of the
relaxation dynamics can be characterized by means of the self-intermediate
scattering function (SISF) $F_s(q,t)$, where $q$ is the wave-vector:

\begin{equation}
F_s(q,t)= \frac{1}{N} \sum_{j=1}^{N} 
\langle \exp [-i \mathbf{q} \cdot (\mathbf{r}_j(t)-\mathbf{r}_j(0))] \rangle\quad .
\label{eq1_fsqt}
\end{equation}

\noindent
Here $N$ is the number of particles considered, $\langle.\rangle$ is
the thermal average, and ${\bf r}_j(t)$ is the position of particle
$j$ at time $t$. In Fig.~\ref{fig1_fsqt} we show the time dependence
of $F_s(q,t)$ for particles that have different connectivity. The
wave-vector is 2.8\AA$^{-1}$ which corresponds to the main peak in the
static structure factor (see Fig.~\ref{figSI_sq} in the SI). We recognize
that for $ t\leq 0.1$~ps the different curves fall on a common curve,
showing that on this time scale the motion of the particles is independent
of their environment. This is no longer the case for somewhat larger
$t$ in that the correlators for large $k$ show a marked peak at around
0.2~ps. The presence of such a peak indicates that on this time scale the
motion of large $k$ particles has a pronounced vibrational character and in the SI we
show that this feature is indeed also very pronounced in the glass state
(see Fig.~\ref{figSI_fsq}). In previous studies the presence of such a
peak has been associated with the so-called boson
peak~\cite{Ang95,Hor96,Kob97,Hor01,Sas03}, i.e.~an excess
in the vibrational density of states in the glass state~\cite{Bin11}. For
the present system this interpretation might not be valid since we
do not find clear evidence that our system has a boson peak (see SI).
Also included in the graph is the correlator for the Cu atoms that are
not in the center of a icosahedron and we see that this function decays
significantly faster than the one for the icosahedrally packed atoms (also
shown), in agreement with the results from Ref.~\cite{li_17}. Furthermore
we see that the former correlator shows no peak at intermediate times,
thus confirming that this feature is related to the highly structured
local environment of the central atom. The upper inset in the figure
shows the same correlators on a much larger time range demonstrating
that the $k-$dependence of the SISF is present even on the time scale of
the $\alpha-$relaxation, see below for more details, thus indicating a
connection between the dynamics at short times with the one
at long times~\cite{Sco03,Luo16,Shi08}.

\begin{figure}[ht]
\centering
\includegraphics[width=1.0\linewidth]{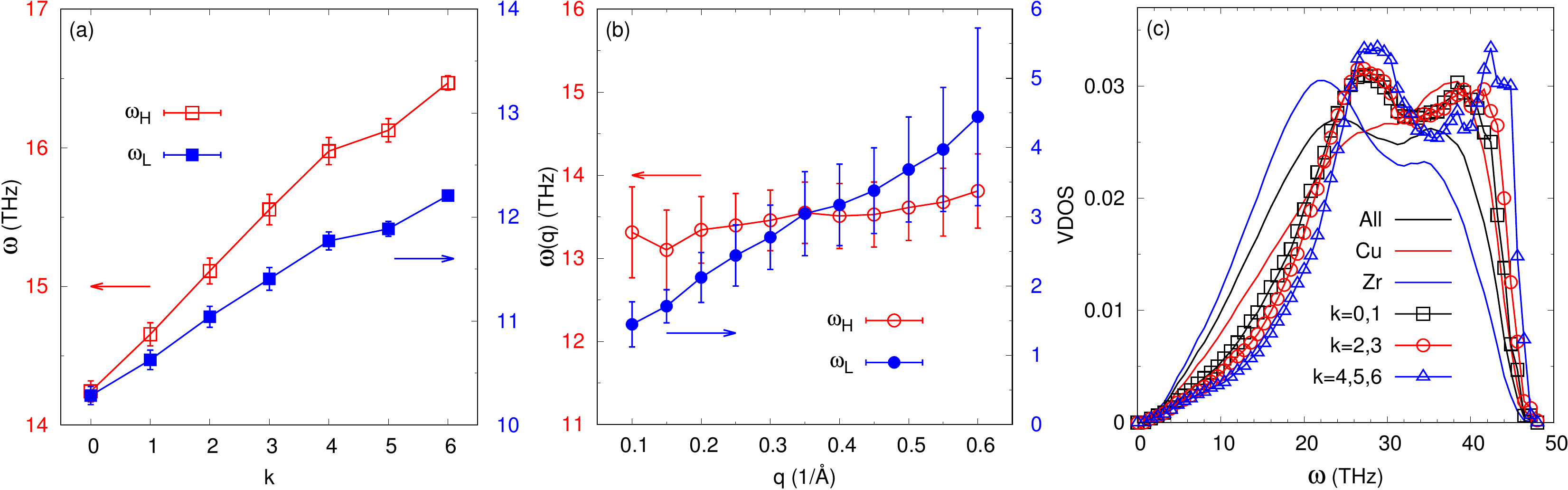}
\caption{{\bf The dependence of the vibrational features on connectivity $k$.}
(a) The frequency of the high and low frequency modes, $\omega_H$ and
$\omega_L$, increases with increasing $k$. (b) The frequency $\omega_H$
is basically independent of the wave-vector $q$ showing
that this is an optical mode whereas the low frequency mode at $\omega_L(q)$
increases linearly with $q$, characteristic of an acoustic mode. 
(c) Vibrational density of states for the different populations showing that the
highly connected icosahedra have on average higher frequencies.
\label{fig2_fitparam}}
\end{figure}

To describe the dynamics at these short times in a 
quantitative manner we fit the correlator with the simple Ansatz

\begin{equation}
F_s(q,t)= \sum_{j=L,H} C_j\exp\{iq[A\cos(\omega_j t+\delta_j)-A\cos(\delta_j)]\},
\label{eq2_fsqfit}
\end{equation}

\noindent
where ``$L$'' and ``$H$'' denote a low and high frequency mode,
respectively and we have $C_L+C_H=1$, i.e.~we describe the motion of a
particle as a superposition of two harmonic oscillators. The resulting
fits are included in Fig.~\ref{fig1_fsqt} as well and we recognize
that this functional form gives indeed a good description of the data.
Figure~\ref{fig2_fitparam}a) presents the $k-$dependence of
$\omega_H$ and $\omega_L$ and one recognizes that both of them increase
(basically linearly) with $k$ which demonstrates that with increasing
connectivity the cage becomes stiffer. Figure~\ref{fig2_fitparam}b)
shows the $q-$dependence of the two frequencies at low $q$ and one sees that
$\omega_L$ is a linear function of $q$, as expected for a generic SISF
that couples at low $q$ to the acoustic modes~\cite{Set98}. In contrast to this,
$\omega_H$ is independent of $q$, showing that this mode has an optical
character. Thus we can conclude that the peak seen in Fig.~\ref{fig1_fsqt}
is related to a spatially localized excitation the details of which, in view of the
observed $k-$dependence in Fig.~\ref{fig1_fsqt}, can be directly related
to the local structure. Similar results are found for the coherent
intermediate scattering function.

In order to put $\omega_H$ and $\omega_L$ in relation with typical
frequencies in the system we show in Fig.~\ref{fig2_fitparam}c)
its density of states at 0K as obtained for the whole system, the
partial densities of states for the two types of atoms, as well as
for the Cu atoms having different connectivities (see Methods). One
sees that with increasing $k$ the peaks at around 28~THz and 42~THz
increase significantly, that the upper limit of the DOS moves to higher
frequencies, and that the DOS between 10 and 20~THz is suppressed. Thus
this $k-$dependence of the DOS is coherent with the result from
Fig.~\ref{fig2_fitparam}a) that the stiffness of the cage increases with
increasing connectivity of the local structure.

\begin{figure}[ht]
\centering
\includegraphics[width=0.8\linewidth]{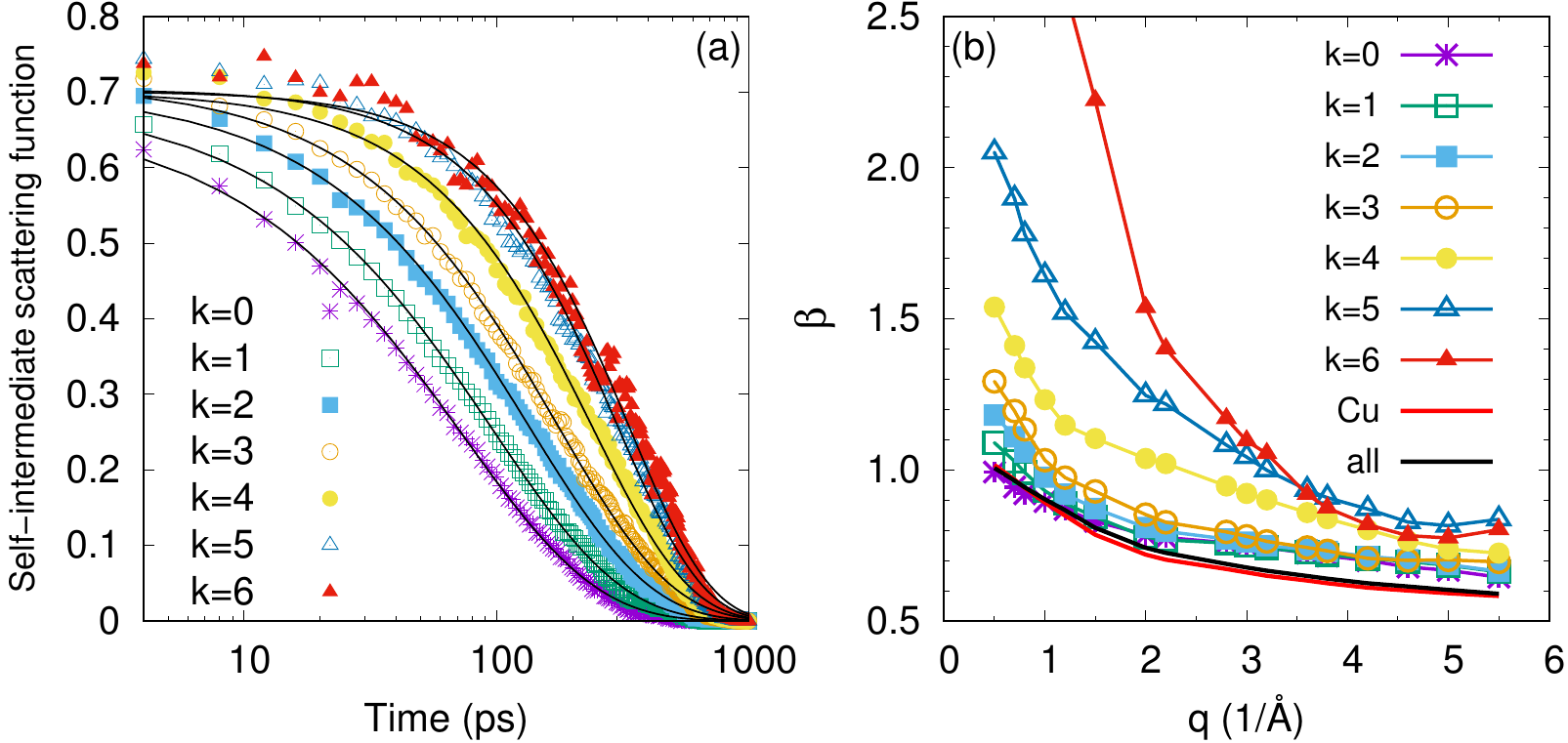}
\caption{{\bf Relaxation dynamics of the system at intermediate and long
times at $T=1000$K.} (a) Self intermediate scattering function at wave-vector
$q=2.8$\AA$^{-1}$ for different values of connectivity $k$ of the Cu
atoms. (b) $q-$dependence of the KWW exponent $\beta$ for different
values of $k$. Also included is the data obtained if one averages over all Cu atoms
and all atoms.}
\label{fig3_fsqt_kww}
\end{figure}

To connect this vibrational motion at short times with the one at
large $t$ we have fitted the SISF for the populations with different
$k$ with a Kohlrausch-Williams-Watts (KWW) function, i.e.~$F_s(q,t)= A
\exp(-(t/\tau)^\beta)$, and Fig.~\ref{fig3_fsqt_kww}a displays this data
and the fits and one sees that for large $k$ the correlator has in the $\alpha-$regime 
a very strong time dependence. Figure~\ref{fig3_fsqt_kww}b presents the $q-$dependence of
the KWW exponent $\beta$ for the different values of $k$. One recognizes that for
small $k$, i.e.~isolated icosahedral clusters, $\beta$ is smaller than 1.0
for all wave-vectors considered, i.e.~the relaxation is stretched as one
expects for a glass-forming system~\cite{Bin11}.  Interestingly we find
that for intermediate and small $q$ the exponent increases significantly
with $k$ and becomes larger than 1.0, i.e.~one sees a crossover from a
normal glassy dynamics to one in which the correlator has a much sharper
decay in time, and that this crossover depends on $q$, i.e.~the length
scale considered. (In the SI we show that with decreasing temperature
this crossover occurs at smaller and smaller $q$'s, i.e.~the length scale
increases.)  Such a quick decay of the correlation function indicates
a sudden yielding of the structure, i.e.~a type of motion that is very
different from the viscous flow found in glassy systems.

\begin{figure}[ht]
\centering
\includegraphics[width=1.0\linewidth]{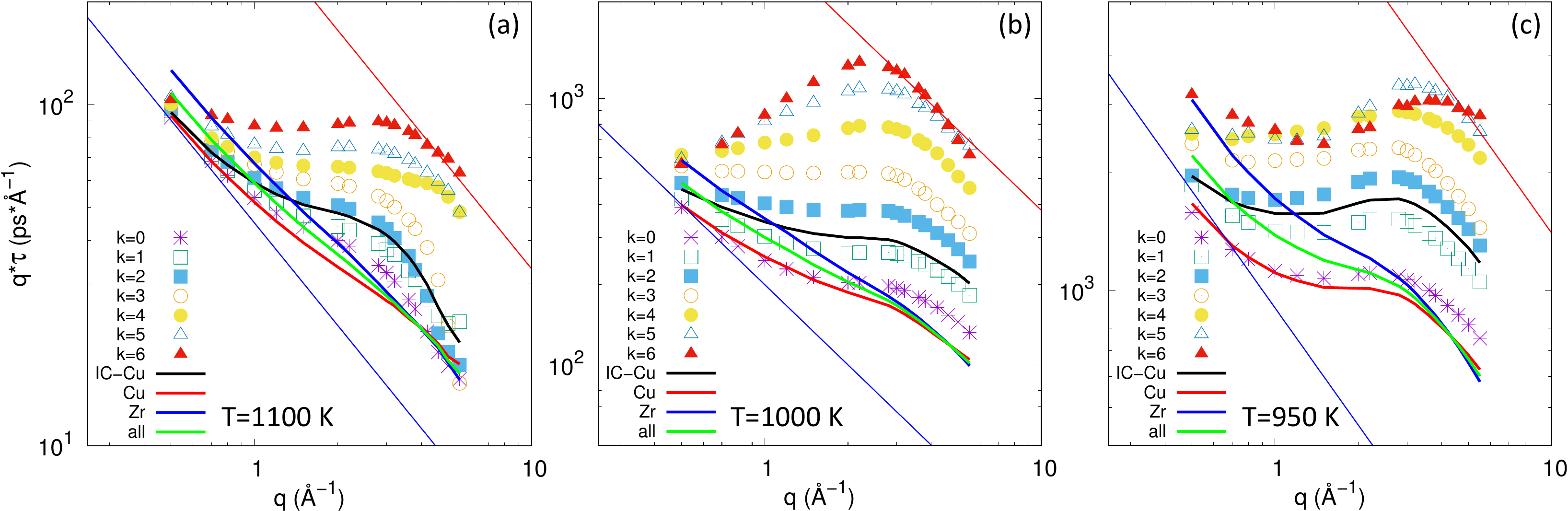}
\caption{{\bf Wave-vector dependence of the $\alpha-$relaxation time.} We
plot $\tau(q) q$ as a function of $q$ for different values of the
connectivity $k$. Also included is the data for all Cu/Zr atoms and all atoms.
The three panels correspond to different temperatures:
(a) $T$=1100K, (b) $T$=1000K, and (c) $T$=950K. The solid lines
are power-laws with exponent $-1$, the $q$ dependence expected for a
diffusive process. 
}
\label{fig4_tau}
\end{figure}

To understand the nature of this abrupt yielding we look at the
$q-$dependence of the $\alpha-$relaxation time $\tau$ as obtained from
the fit with the KWW form. For a diffusive motion one expects that
$\tau \propto q^{-2}$ whereas for a ballistic displacement one has $\tau
\propto q^{-1}$. To allow for a better distinction of these two cases we
plot in Fig.~\ref{fig4_tau} the product $\tau(q) \cdot q$ as a function
of $q$. The data shows that at high $T$, panel (a), the $q-$dependence
of $\tau$ is close to $q^{-2}$ if $k$ is small, i.e.~the dynamics is
diffusive, as expected for a supercooled liquid at intermediate and small
$q$. For large $k$ we find, however, a plateau in the $q-$dependence,
indicating that on these length scales the relaxation dynamics is
ballistic, i.e.~the icosahedron moves like a rigid structure. This
shows that icosahedra which have in their first nearest neighbor shell
many Cu atoms that are themselves at the center of an icosahedron are
sufficiently rigid to be convected by the surrounding medium.

If $T$ is lowered to 1000K, panel~(b), we find that now also the curves
for small and intermediate $k$ show a plateau, i.e.~at this temperature
also these icosahedra have become sufficiently rigid to be moved as a
rigid block. For the icosahedra with large $k$ one recognizes that $\tau$
decays somewhat slower than $q^{-1}$ which implies a relaxation that is
faster than a rigid translation dynamics. We see that for the largest
$k$'s this rapid relaxation dynamics is observed in a range $0.5 {\rm
\AA}^{-1} \leq q \leq 2 {\rm \AA}^{-1}$, i.e.~it extents from about
one interparticle distance up to distances that are on the order of
the size of a few icosahedra, thus about one nm. From the fact that
on these length scales the relaxation time is almost constant we can
conclude that the underlying structure breaks up in a very abrupt manner,
indicating a yielding process that is more similar to the one found
in a solid that fractures than to the continuous transformation found
in a liquid. Hence this suggests that high $k$ icosahedra form rigid
structures that are moved by the surrounding viscous medium which in turn is driven
by local internal stresses.

Comparing this data with the one at $T=1100$K shows that the decrease of
$T$ affects more strongly the relaxation dynamics for the icosahedra with
high $k$ than the ones with low $k$. This implies that for this system
the so-called $\alpha-$scale universality predicted by mode-coupling
theory~\cite{gotze_book_08} does not hold, i.e.~the relaxation time
$\tau_x(T)$ for an arbitrary observable $x$ cannot be written as
$\tau_x(T)=h_x f(T)$, where $h_x$ is a $x-$dependent constant and $f(T)$
is a system universal function of temperature. (In our case $x$ can be
$q$ or $k$.)  The fact that the $\alpha-$scale universality does not
hold for the present system, whereas it works for simple glass-forming
systems such as binary Lennard-Jones mixtures~\cite{gleim_00}, shows that
metallic glass-formers have a surprisingly complex local dynamics and
hence mean-field-like theories are no able to catch these features of
the relaxation.

If temperature is lowered even more, see data for $T=950$K in panel (c),
the unusual $q-$dependence found for large $k$ is less pronounced and the
relaxation dynamics becomes again more or less ballistic. At the same time the
curves for low $k$ show a plateau that is more pronounced than the one
at higher $T$s which makes that on overall the $k-$dependence of $\tau$
becomes weaker, i.e.~the relaxation dynamics of the system becomes more
homogeneous. The likely reason for these changes is that a decrease
in temperature will lead to high $k$ icosahedra that are somewhat more
rigid but that the viscosity, which is roughly proportional to $\tau$,
increases significantly. Hence the shear forces acting on the icosahedra
will have increased substantially, making the latter to break up in a
more continuous manner and thus leading to the disappearance of the
faster than ballistic regime in $\tau(q)$. Thus we can conclude that
the different mechanical properties of the icosahedra with different $k$
make that the nature of their relaxation dynamics depends strongly on $k$
as well as temperature.

Also included in the graphs are the relaxation times as obtained for all
the Cu atoms that are in the center of an icosahedron (black solid line)
and one sees that the discussed change in transport mechanism makes that
the $q-$dependence of $\tau$ for this population leads to the formation
of a small shoulder at 1000K and a small hump at 950K. If {\it all} Cu
atoms are considered (red line) one finds at high and intermediate
temperatures only a weak shoulder and at the lowest temperature a
pronounced plateau. In contrast to this the Zr atoms have at high and
intermediate temperatures a $q-$dependence that is close to the one for
a diffusive dynamics and only at the lowest temperature one can notice a
weak shoulder (blue line). The same is true if one looks at the $\tau(q)\cdot q$
when averaged over all atoms (solid green line). These results thus
demonstrate that the described anomalous behavior in the relaxation
dynamics is detectable also in real experiments that can only measure
the system averaged quantities.

\begin{figure}[ht]
\centering
\includegraphics[width=0.9\linewidth]{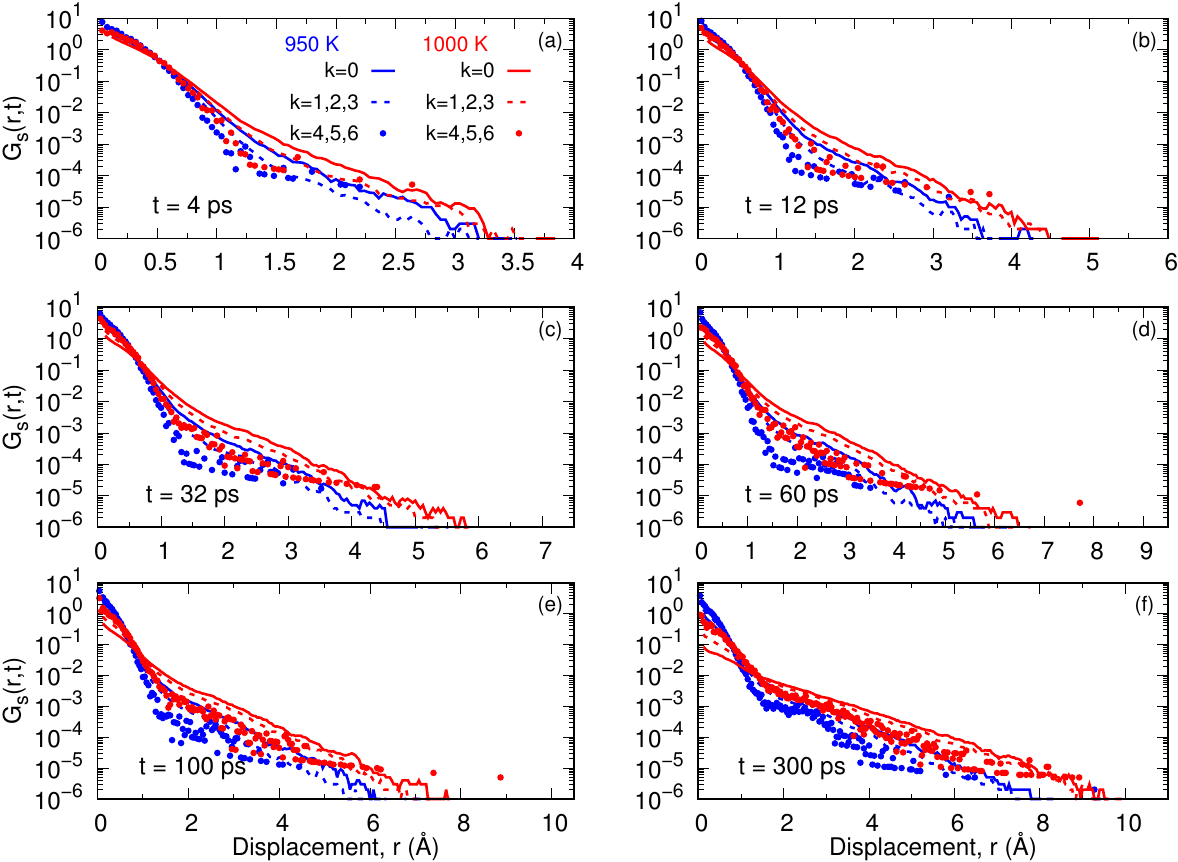}
\caption{
{\bf Space dependence of the self part of the van Hove correlation function
$G_s(r,t)$ for different types of icosahedral clusters:} The curves
correspond to $k=0$ (full lines), the average over $k=1,2,3$ (dashed
lines), and the average over $k=4,5,6$ (dotted lines). The red and blue
curves are for $T=1000$K and $T=950$K, respectively. The panels correspond
to different times $t$ (given in the panels). 
}
\label{fig5_gsrt}
\end{figure}

So far we have discussed the relaxation dynamics of the particles in
reciprocal space. Since in simulations one can also easily access the
trajectories of the particles in real space we can use this information
to check the interpretation of the $q-$space data. For this we have 
determined the self part of the van Hove function defined as~\cite{Han06}

\begin{equation}
G_{\rm s}(r,t)= \frac{1}{N} \sum_{i=1}^{N}
\left\langle  \delta(r-|{\bf r}_i(0)-{\bf r}_i(t)|) \right\rangle .
\label{eq_vanhove}
\end{equation}

\noindent
This distribution is shown in Fig.~\ref{fig5_gsrt} for different times
$t$ and values of $k$. One recognizes that for $k=0$ it is given by a
Gaussian at short distances and by an exponential tail at large $r$,
i.e. the typical behavior found in glassy liquids~\cite{chaudhuri_07}. For
$k>0$ we find that the distribution at distances larger than $r\gtrsim
1.0$\AA~and intermediate times has a much smaller slope than the one
for $k=0$ which shows that particles with high $k$ explore these
large distances on the same time scale, in agreement with the results shown in
Fig.~\ref{fig4_tau}. For large times, panel f), the distribution for the
different $k$ become similar to each other which is likely connected to
the fact that the atoms have changed their $k$ values and hence loose
their memory regarding the nature of their neighborhood at $t=0$. Since
the mean squared displacement (MSD) is just the second moment of
$G_s(r,t)$ this $k-$dependence is also seen in the time dependence of
the MSD in that it shows for large $k$ a jump-like increase (see SI),
a $t-$dependence that is again very different from the one found in
usual glass-forming liquids~\cite{Bin11}.

Our results show that even in the fluid state certain glassy systems can
show a relaxation dynamics that is given by a compressed exponential,
i.e.~a time dependence that differs strongly from the usual stretched
exponential found in viscous liquids. For the metallic glass-former
investigated here these two type of relaxation dynamics do even
coexist and can be directly related to the local atomic structure. The
relative importance of the two relaxation mechanisms depends strongly
on temperature as well as the length scale considered, i.e.~the
wave-vector,and becomes very pronounced at around the mode-coupling
temperature of the system, and gives rise to a decoupling phenomena in the
dynamics that is non-monotonic in temperature. It can be expected that
the structural features that give rise to these different relaxation
mechanisms in the liquid are also responsible for the interesting
mechanical properties of metallic glasses, such as elasticity and
ductility, in that the more rigid structures formed by the high $k$
icosahedra are embedded by a softer surrounding that allows for localized
plastic relaxation.  Thus our findings show that a careful investigation
of the wave-vector dependence of the relaxation dynamics allows to get
deeper insight into the unique properties of metallic glasses, insight
that will help to conceive other new materials that do have attractive
mechanical properties.\\[20mm]

{\bf Methods}
Simulations have been carried out using the LAMMPS software~\cite{Pli95}.
The time step was 1~fs and the initial atomic configuration was firstly equilibrated
for long time (2~ns) at temperature T= 2000~K in the NPT ensemble ($P= 0$~bar) using a
Nose-Hoover thermostat and barostat. The liquid was then cooled down to
its target temperature at rate of 1 K/ps at constant pressure $P= 0$~bar and
subsequently relaxed for 1~ns before structure and dynamics data were collected.

The vibrational density of states was calculated by making a run at
low temperatures and calculating the time Fourier transform of the
velocity-autocorrelation function of the species of interest, i.e.~Cu, Zr,
and Cu atoms with different $k$. We have checked that this approach gives
the same results as a direct diagonalization of the dynamical matrix.
See SI for a comparison between the two methods.

\begin{acknowledgments}
We thank B. Ruta, M. Z. Li, H. P. Zhang, P. F. Guan, and R. Blumenfeld for useful
discussion. This work is supported by the National Basic Research
Program of China (973 Program, Grant No. 2015CB856801),
National Natural Science Foundation of China (Grant No. 11525520),
and China Postdoctoral Science Foundation (Grant No. 2017M610687).
\end{acknowledgments}

\vspace*{10mm}

%
%
%
%
%
%
%

\renewcommand{\thefigure}{S\arabic{figure}}
\setcounter{figure}{0}

\newpage
{\Large {\bf Supplemental Material}}

\begin{figure}[ht]
\centering
\includegraphics[width=1.0\linewidth]{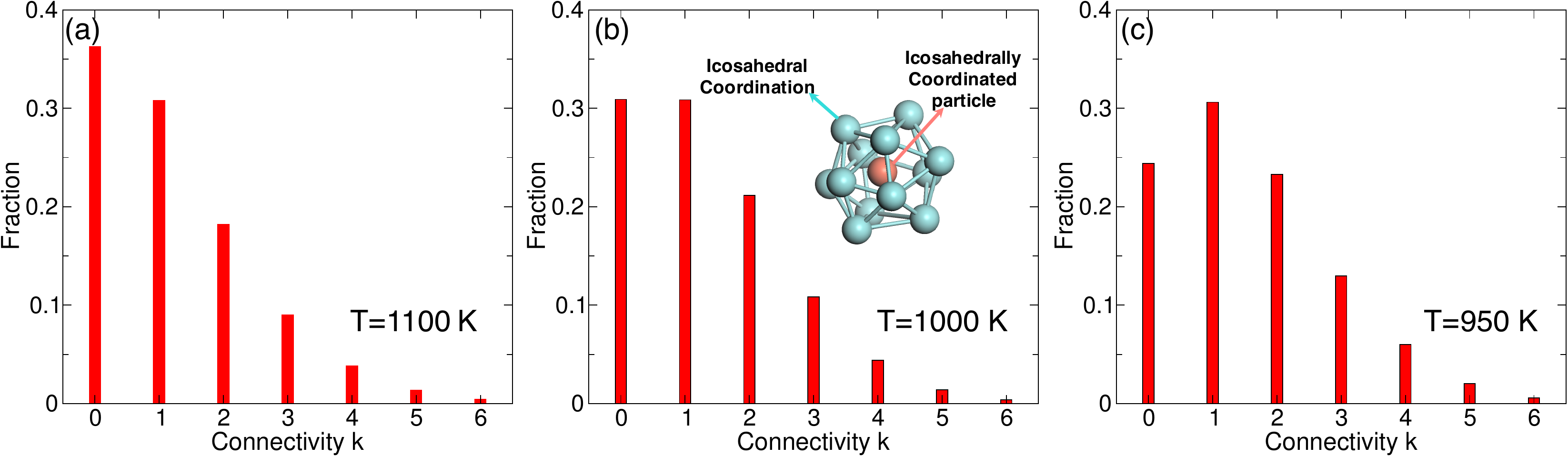}
\caption{
Probability that a icosahedron is of type $k$. (a) $T=1100$~K, (b)
$T=1000$~K, (c) $T=950$~K. The inset in panel (b) shows a snapshot 
with a Cu at the center of an icosahedron.
}
\label{figSI_ico}
\end{figure}

\begin{figure}[bht]
\centering
\includegraphics[width=0.45\linewidth]{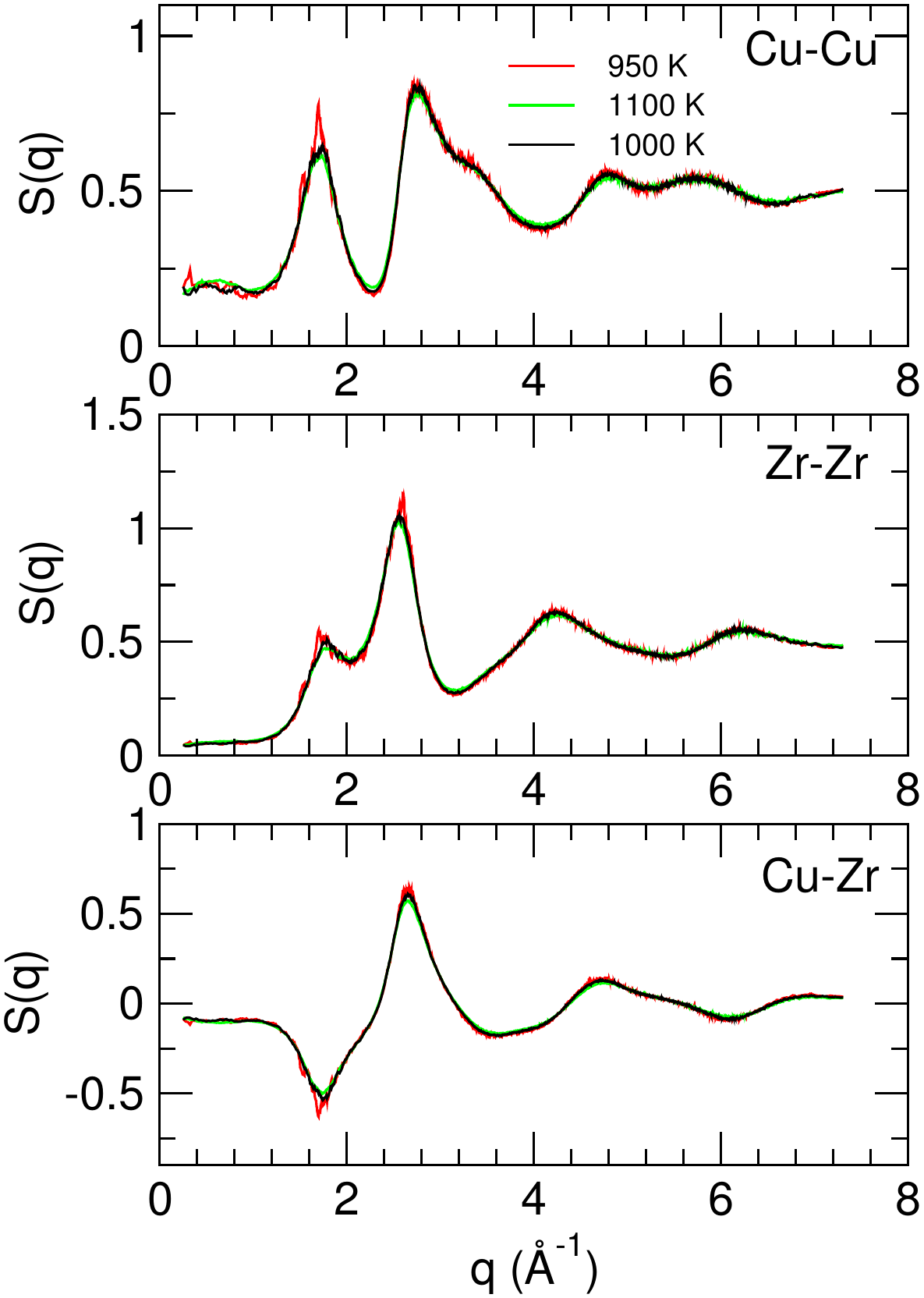}
\caption{
$q-$dependence of the partial structure factors for the three temperatures
considered.
}
\label{figSI_sq}
\end{figure}

In Fig.~\ref{figSI_ico} we show the probability that an icosahedron
with a Cu atom at its center has connectivity $k$.  We see that at high
temperatures most icosahedra are of type $k=0$ and that with decreasing
temperature the icosahedra become increasingly connected.

Figure~\ref{figSI_sq} shows the $q-$dependence of the partial static
structure factors at three different temperatures. One recognizes that
these functions show basically no $T-$dependence. The main peak for the
Cu-Cu correlation is around 2.8\AA$^{-1}$, the wave-vector we often
focus on in the present study.\\[-5mm]

\clearpage

\begin{figure}[t] \centering
\includegraphics[width=0.6\linewidth]{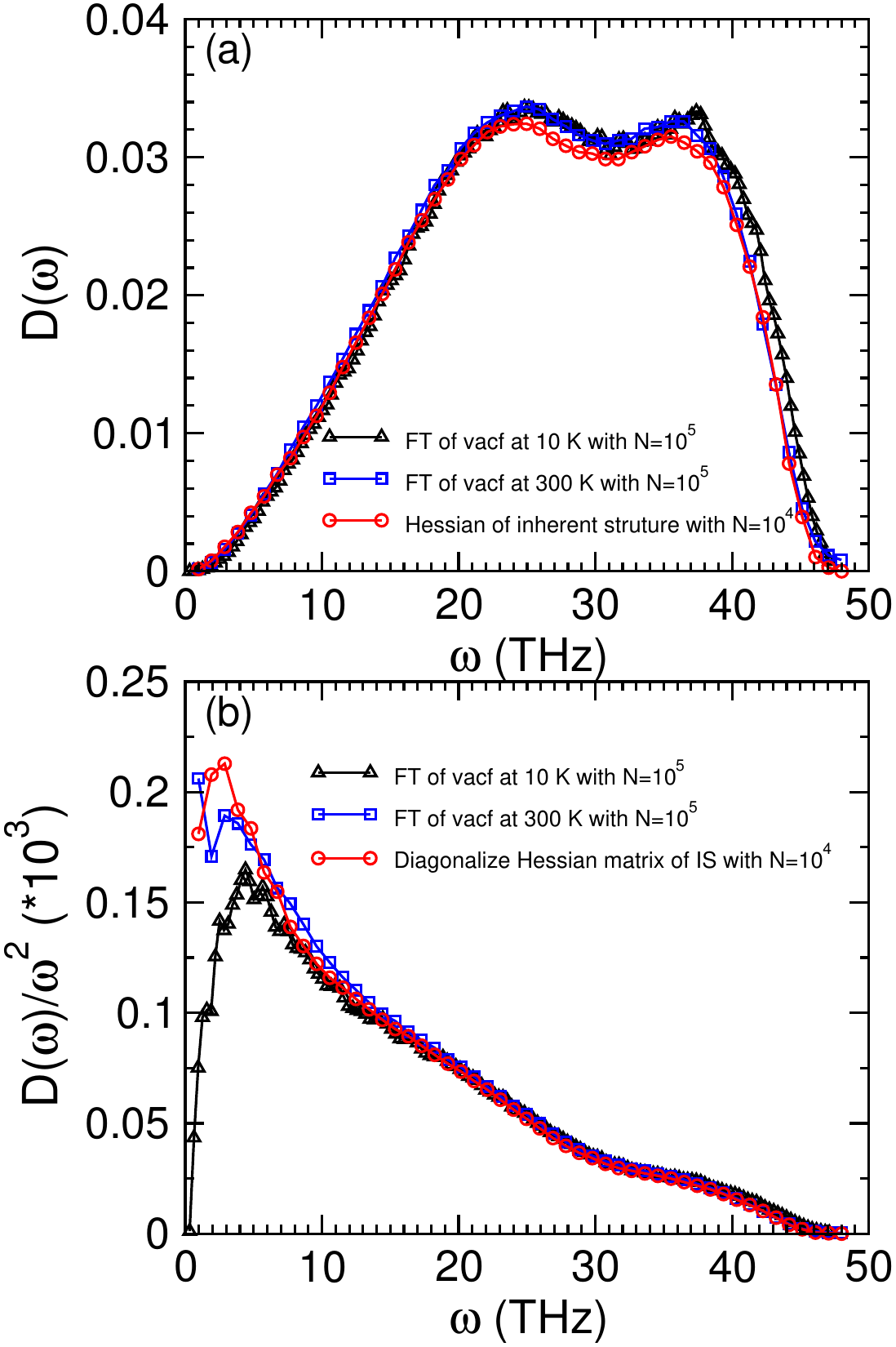} 
\caption{ a) Vibrational density of states for $N=10^4$ particles and
$N=10^5$ particles.  For the system with $N=10^4$ particles we have
obtained the VDOS by the direct diagonalization of the Hessian matrix and
for the $N=10^5$ system by the calculation of the time Fourier transform
of the velocity auto-correlation function.  b) VDOS/$\omega^2$ to see
whether or not there is a boson peak.} 
\label{figSI_vdos}
\end{figure}

Figure~\ref{figSI_vdos}a) shows the vibrational density of states (VDOS)
$D(\omega)$ for two system sizes. For $N=10^5$ particles we have computed
$D(\omega)$ by calculating the time-Fourier transform of the velocity
autocorrelation function of the particles. The data for $T=300$K agrees
very well with the one obtained at $T=10$K, indicating that we are
indeed probing the harmonic regime. For the smaller system size with $N=10^4$
particles we have obtained $D(\omega)$ by diagonalizing directly the
Hessian matrix evaluated at a local minimum of the potential energy of
the system. The so obtained VDOS agrees well with the one
obtained for the larger system.

In Fig.~\ref{figSI_vdos}b) we show the vibrational density of states
divided by $\omega^2$. This type of plot can be used to probe for the
existence of a boson peak in the system. The data shows that there
is no strong evidence for the presence of a peak at low frequencies,
i.e.~no marked boson peak, although the existence of such an excitation
at around 3THz cannot be excluded.

\begin{figure}[ht]
\centering
\includegraphics[width=0.6\linewidth]{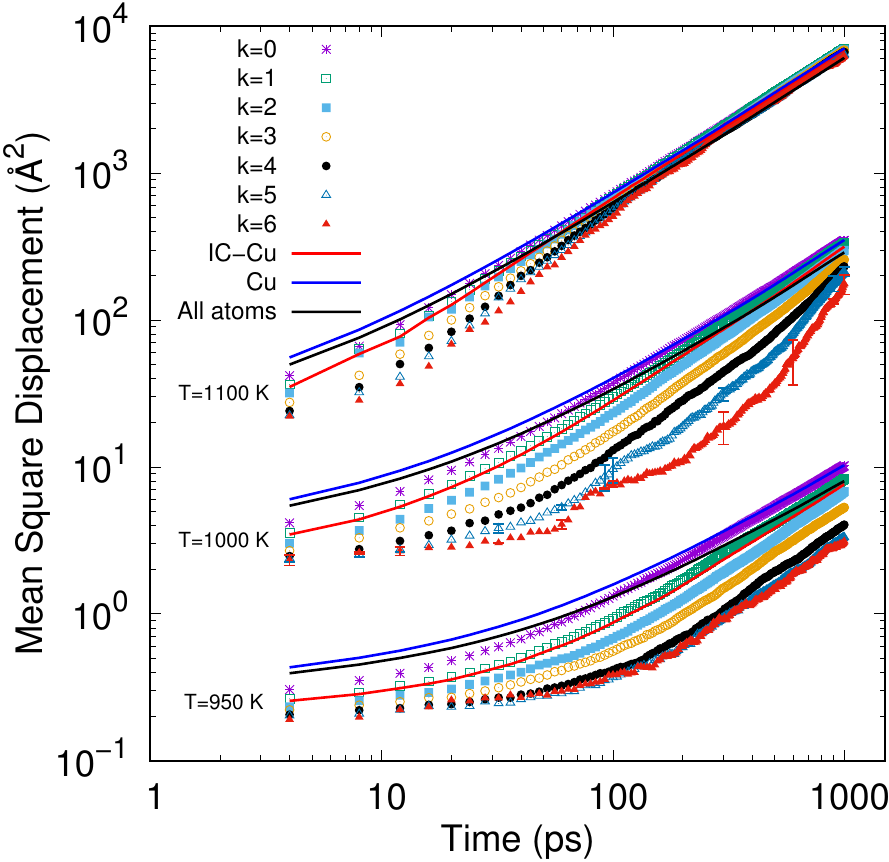}
\caption{
Mean squared displacement for different of different type of atoms for 
three different temperatures. Cu: Any Cu atoms; IC-Cu: Any Cu that is at the
center of an icosahedron; the data labeled with $k$ are the MSD for Cu
with connectivity $k$.  For $T=1100$~K and $T=1000$~K the curves have
been shifted upward by a factor of 10 and 100, respectively.
}
\label{figSI_msd}
\end{figure}

\clearpage
In Fig.~\ref{figSI_msd} we show the time dependence of the mean squared
displacement (MSD) of various species: Cu atoms, Cu atoms inside an
icosahedron, Cu atoms inside an icosahedron of connectivity $k$, and
average over all atoms. One sees that the IC-Cu atoms move slower that
the average Cu atom which in turns move a bit faster than the Zr atoms,
results that are in agreement with previous studies.  For $T=1100$K
the $k-$dependence of the MSD is relatively weak and all the curves are
smooth. For $T=1000$K there is a more pronounced $k-$dependence of the
MSD and the data for large $k$ shows a rapid variation on the time scale
of 50-200ps. These steps are an indication that in this time regime the
particles undergo a rapid motion, compatible with the view expressed in
the main text that these high $k$ structures break up in a very sudden
manner. If $T$ is lowered to 950K, these steps are no longer visible,
in agreement with the discussion in the main text regarding the shape
of the time correlation functions.

\clearpage

\begin{figure}[h]
\centering
\includegraphics[width=0.8\linewidth]{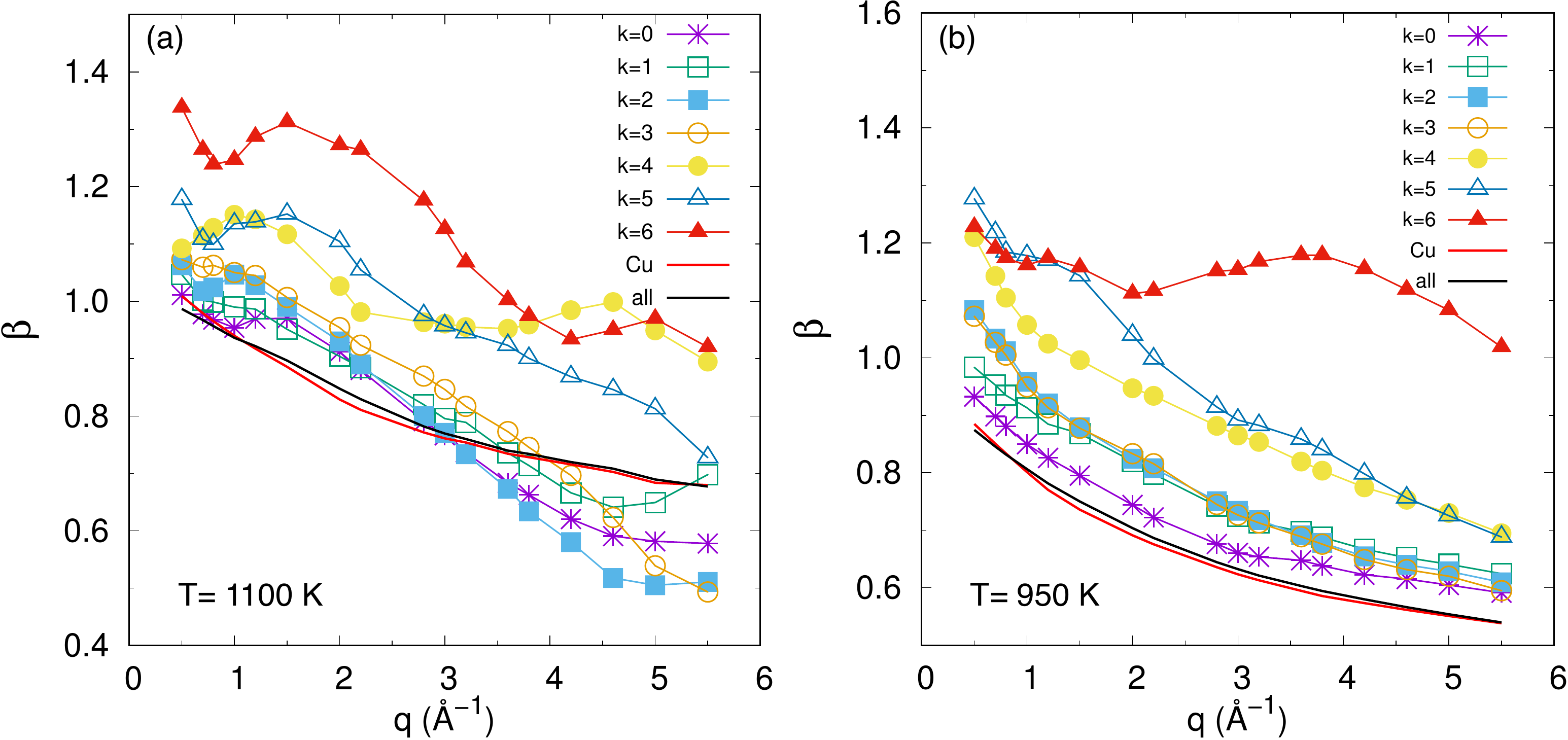}
\caption{
$q-$dependence of the KWW exponent $\beta$ for different connectivity $k$ for $T=1100$K 
and $T=950$K, panels a), and b), respectively.
}
\label{figSI_betaq}
\end{figure}

In Fig.~3b) of the main text we have shown the wave-vector dependence of
the stretching parameter $\beta$ for $T=1000$K. In Fig.~\ref{figSI_betaq}
we show this dependence for $T=1100$K and $T=950$K. A comparison of
these three figures shows that at the intermediate temperature the
$q-$dependence for the large $k$ clusters is much stronger than at the
lower and higher $T$. For low $k$ clusters this dependence is instead
only a weak function of temperature. Thus these results support the
conclusion from the main text that for temperature around 1000K there is
a strong dependence of the nature of the relaxation dynamics for large
$k$ clusters.

\clearpage

\begin{figure}[ht]
\centering
\includegraphics[width=0.6\linewidth]{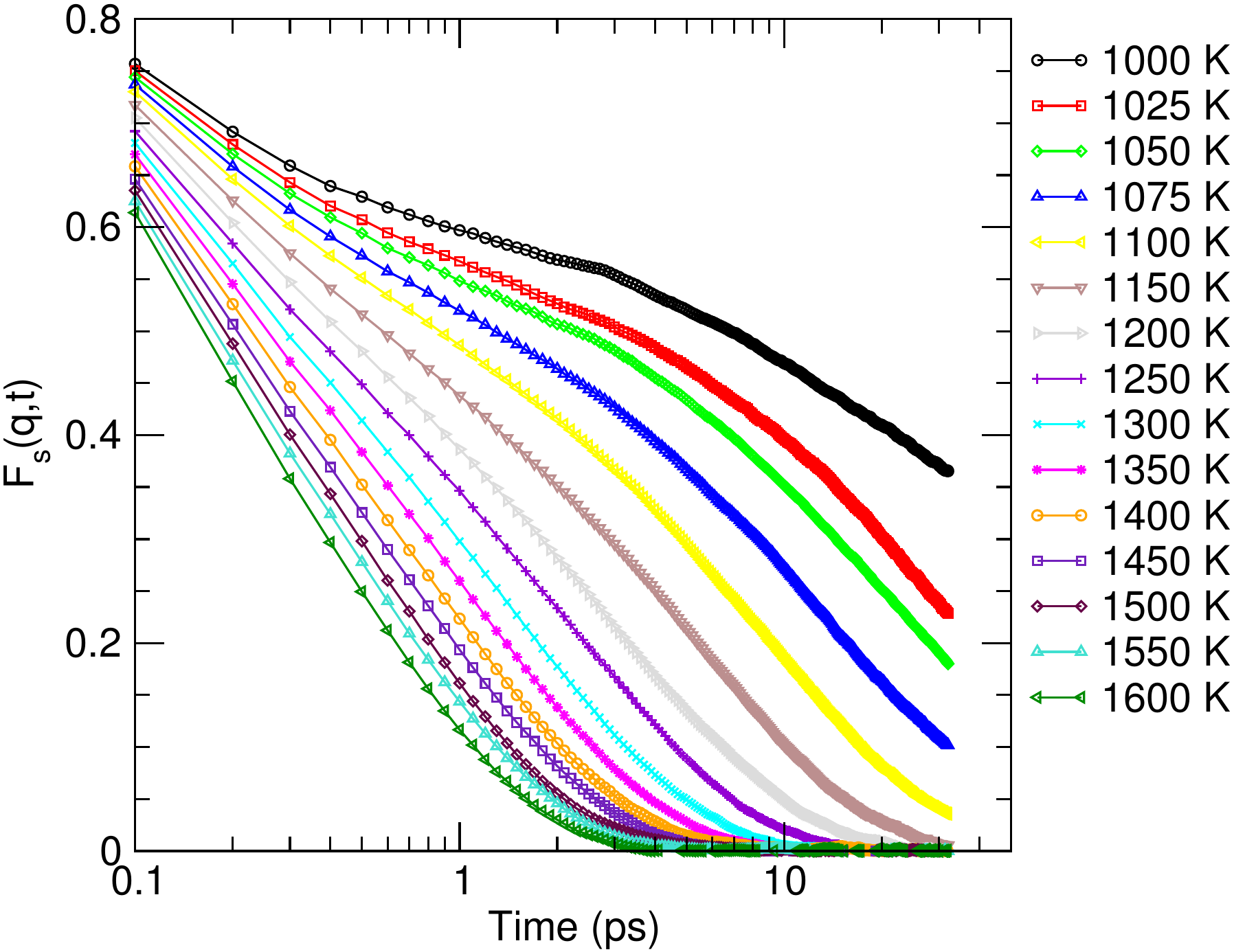}
\caption{
Time dependence of the self intermediate scattering function for the Cu
atoms at different temperatures.  The wave-vector is $q=2.8$\AA$^{-1}$.
}
\label{figSI_Fs_diffT}
\end{figure}

In Fig.~\ref{figSI_Fs_diffT} we show the time dependence of the
self intermediate scattering function of the Cu atoms for different
temperatures. The wave-vector is 2.8\AA$^{-1}$.  We see that for $T$
around 1250K the correlator starts to show a weak shoulder and thus we
can identify this with the onset temperature of the system.

\begin{figure}[ht]
\centering
\includegraphics[width=0.4\linewidth]{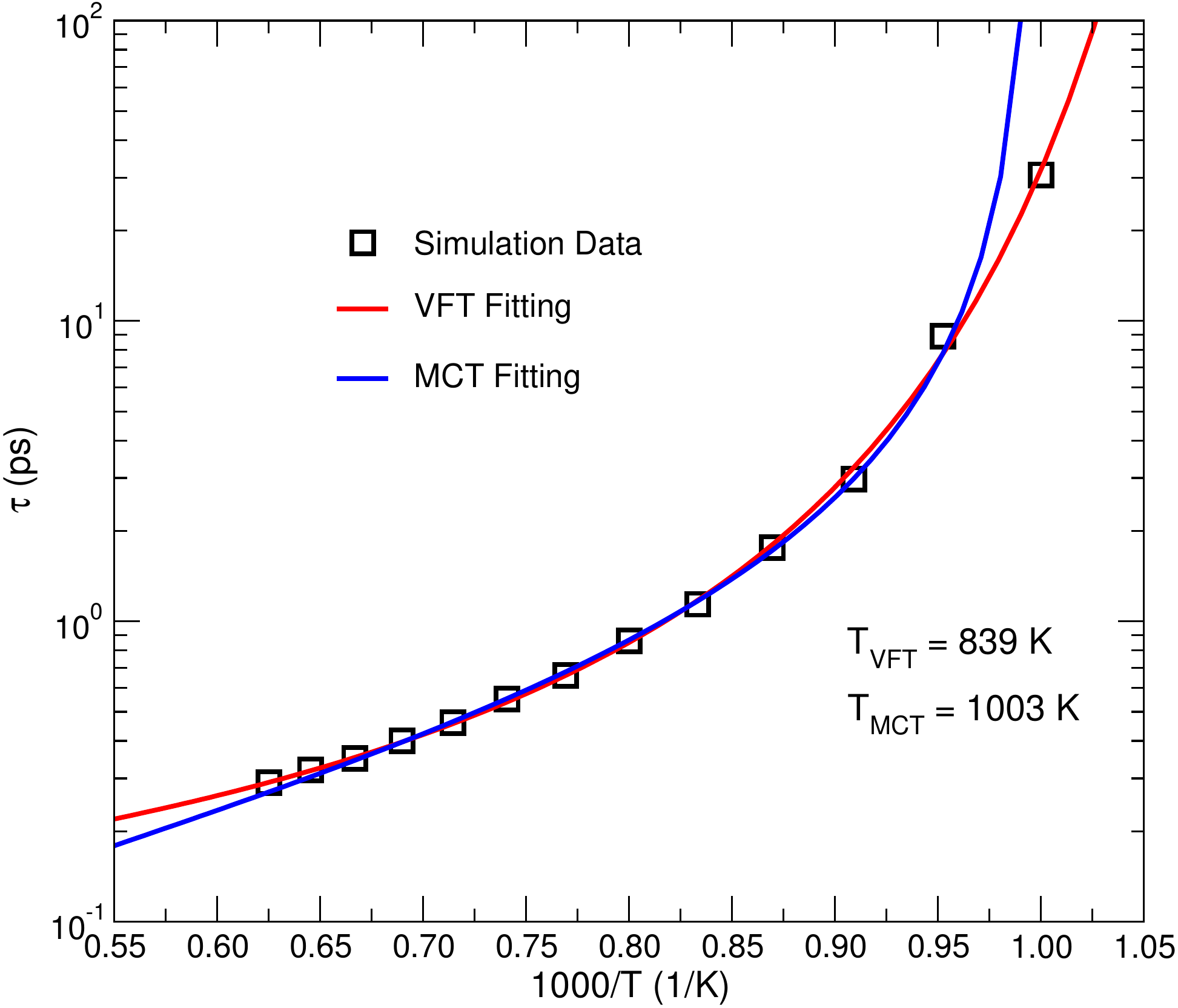}
\caption{
Arrhenius plot of the relaxation time as obtained from the self intermediate scattering function
for the Cu atoms at wave-vector $q=2.8$\AA$^{-1}$. 
}
\label{figSI_tau_arrhenius}
\end{figure}

\begin{figure}[ht]
\centering
\includegraphics[width=0.6\linewidth]{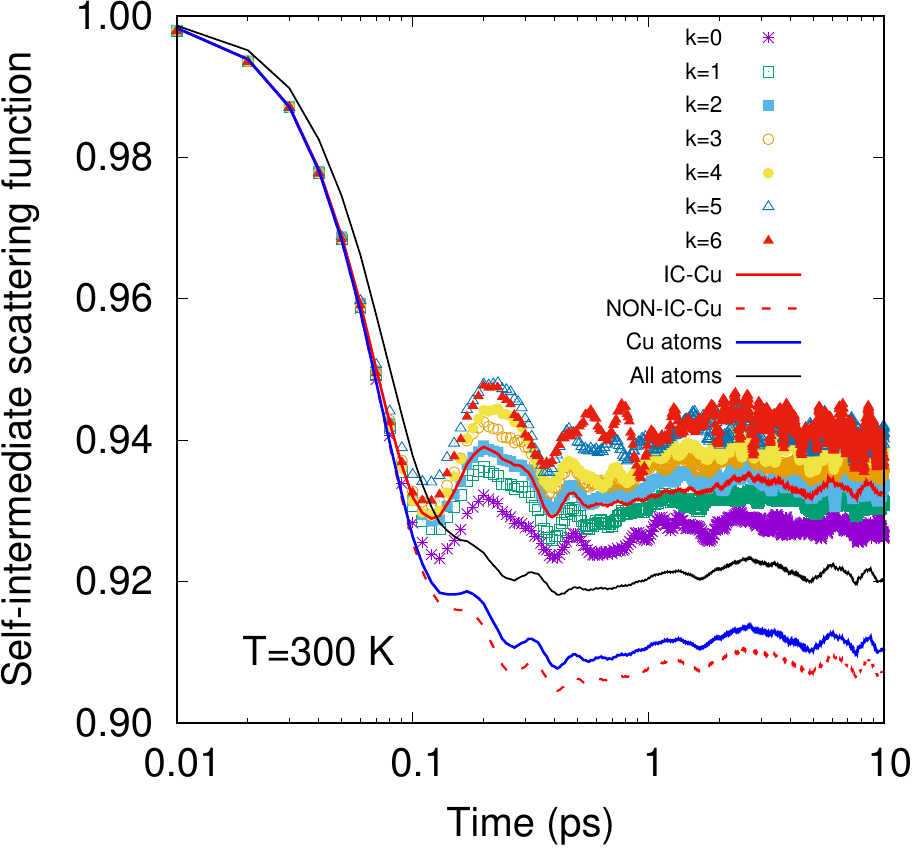}
\caption{
Self intermediate scattering function for $q=2.8$\AA$^{-1}$
at $T=300$K. The different curves correspond to different values of
$k$. Also included is the data for all the Cu atoms and for all atoms.
}
\label{figSI_fsq}
\end{figure}

From the correlation functions shown in Fig.~\ref{figSI_Fs_diffT} one
can obtain the $\alpha$-relaxation time $\tau(T)$ via the definition
$F_s(q,\tau)=e^{-1}$. Figure~\ref{figSI_tau_arrhenius} is an Arrhenius
plot of $\tau$ and it demonstrates that the system has as quite strong
non-Arrhenius behavior. Also included in the graph are fits to the data
with a Vogel-Fulcher-Tammann law, $\tau(T) \propto \exp[E/(T-T_{\rm
VFT})]$, as well as a fit to the data with the power-law proposed by
mode-coupling theory, $\tau \propto (T-T_{\rm MCT})^\gamma$. At this
stage we do not want to put much emphasis on the quality/significance
of these fitting functions but we include them in order to obtain a
better idea on the relevant temperature scales of the system. From
these temperatures we thus can conclude that the change in transport
mechanism discussed in the main text occurs at a temperature that is
close to $T_{\rm MCT}$, the critical temperature of mode-coupling theory.

\clearpage 

The results discussed in the main text concern mainly the dynamics of
the system in its supercooled liquid state. However, one expects that
the $k-$dependence of the vibrational features on the sub-picosecond
time scale can also be observed in the glass-state. That this is indeed
the case is demonstrated in Fig.~\ref{figSI_fsq} where we show the
self intermediate scattering function at 300K for various types of
particles. One recognizes that the Cu atoms with high $k$ do indeed
show a vibrational motion that is much more pronounced than the one for
the Cu with low $k$, in agreement with expectation. This result is also
coherent with the $k-$dependence of the vibrational density of states
shown in Fig.~3c) of the main text.

\begin{thebibliography}{}


\bibitem{Deb01} P. G. Debenedetti and F. H. Stillinger, Nature {\bf 410}, 259 (2001).

\bibitem{Bin11} K. Binder and W. Kob, \emph{Glassy Materials and Disordered Solids:
An Introduction to Their Statistical Mechanics} (World Scientific, 2011).

\bibitem{varshneya_06}
A. K. Varshneya,
{\it Fundamentals of inorganic glasses, 2nd edition} (Society of Glass Technology, 2006)

\bibitem{Edi96} M. D. Ediger, C. A. Angell, and S. R. Nagel, J. Phys. Chem. {\bf 100}, 13200 (1996).

\bibitem{Kob95} W. Kob and H. C. Andersen, Phys. Rev. E {\bf 52}, 4134 (1995).

\bibitem{Kob99} W. Kob, J. Phys.: Condens. Matter {\bf 11}, R85 (1999).

\bibitem{Puo17} F. Puosi, N. Jakse, and A. Pasturel, 	arXiv:1711.02298 [cond-mat.soft]

\bibitem{berthier_11}
L. Berthier, G. Biroli, J.-P. Bouchaud, L. Cipelletti, and W. van Saarloos (Eds.),
{\it Dynamical heterogeneities in glasses, colloids and granular materials}
(Oxford University Press, Oxford, 2011).

\bibitem{Rut14} B. Ruta, G. Baldi, Y. Chushkin, B. Ruffl$\acute{\rm e}$, L. Cristofolini, A. Fontana, M. Zanatta, and F. Nazzani, Nat. Commun. {\bf 5}, (2014).

\bibitem{Rut12} B. Ruta, Y. Chushkin, G. Monaco, L. Cipelletti, E. Pineda, P. Bruna, V. M. Giordano, and M. Gonzalez-Silveira, Phys. Rev. Lett. {\bf 109}, 165701 (2012).

\bibitem{Cip00} L. Cipelletti, S. Manley, R. C. Ball, and D. A. Weitz, Phys. Rev. Lett. {\bf 84}, 2275 (2000).

\bibitem{Bal08} P. Ballesta, A. Duri, and L. Cipelletti, Nat. Phys. {\bf 4}, 550 (2008).

\bibitem{Car08} C. Caronna, Y. Chushkin, A. Madsen, and A. Cupane, Phys. Rev. Lett. {\bf 100},055702 (2008).

\bibitem{Guo09} H. Guo, G. Bourret, M. K. Corbierre, S. Rucareanu, R. B.Lennox,
K. Laaziri, L. Piche, M. Sutton, J. L. Harden, and R. L. Leheny,
Phys. Rev. Lett. {\bf 102}, 075702 (2009).

\bibitem{ciccotti_09}
M. Ciccotti,
J. Phys. D {\bf 42}, 214006 (2009).

\bibitem{sun_15}
B. A. Sun and W. H. Wang,
Prog. Mat. Science {\bf 74}, 211 (2015).

\bibitem{Luo17} P. Luo, P. Wen, H. Y. Bai, B. Ruta, and W. H. Wang, Phys. Rev. Lett. {\bf 118}, 225901 (2017).

\bibitem{lad_12}
K. N. Lad, N. Jakse, and A. Pasturel,
J. Chem. Phys. {\bf 136}, 104509 (2012).

\bibitem{jaiswal_15}
A. Jaiswal, T. Egami, and Y. Zhang,
Phys. Rev. B {\bf 91}, 134204 (2015).

\bibitem{Wak10} M. Wakeda and Y. Shibutani, Acta Mater {\bf 58}, 3963 (2010).

\bibitem{Wu13a} Z. W. Wu, M. Z. Li, W. H. Wang, and K. X. Liu, Phys. Rev. B {\bf 88}, 054202 (2013).

\bibitem{Li09} M. Li, C. Z. Wang, S. G. Hao, M. J. Kramer, and K. M. Ho, Phys. Rev. B {\bf 80}, 184201 (2009).

\bibitem{li_17}
F. X. Li and M. Z. Li,
J. Appl. Phys. {\bf 122}, 225103 (2017).

\bibitem{She06} H. W. Sheng, W. K. Luo, F. M. Alamgir, J. M. Bai, and E. Ma, Nature {\bf 439}, 419 (2006).

\bibitem{Men07} M. I. Mendelev, D. J. Sordelet, and M. J. Kramer, J. Appl. Phys. {\bf 102}, 043501 (2007).

\bibitem{Fin77} J. L. Finney, Nature {\bf 266}, 309 (1977).

\bibitem{Wu16} Z. W. Wu, F. X. Li, C. W. Huo, M. Z. Li, W. H. Wang, and K. X. Liu, Sci. Rep. {\bf 6}, 35967 (2016).

\bibitem{Ang95} C. A. Angell, Science {\bf 267}, 1924 (1995). 

\bibitem{Hor96} J. Horbach, W. Kob, K. Binder, and C. A. Angell, Phys. Rev. E {\bf 54}, R5897 (1996).

\bibitem{Kob97} W. Kob and J.-L. Barrat, Phys. Rev. Lett. {\bf 78}, 4581 (1997).

\bibitem{Hor01} J. Horbach, W. Kob, and K. Binder, Eur. Phys. J. B {\bf 19}, 531 (2001).

\bibitem{Sas03} S. Sastry and C. Austen Angell, Nat. Mater. {\bf 2}, 739 (2003).

\bibitem{Sco03} T. Scopigno, G. Ruocco, F. Sette, and G. Monaco, Science {\bf 302}, 849 (2003).

\bibitem{Luo16} P. Luo, Y. Z. Li, H. Y. Bai, P. Wen, and W. H. Wang, Phys. Rev. Lett. {\bf 116}, 175901 (2016).

\bibitem{Shi08} H. Shintani and H. Tanaka, Nat. Mater. {\bf 7}, 870 (2008).

\bibitem{Set98} F. Sette, M. H. Krisch, C. Masciovecchio, G. Ruocco, and G. Monaco, Science {\bf 280}, 1550 (1998).

\bibitem{gotze_book_08}
W. G\"otze, {\it Complex dynamics of glass-forming liquids:
A mode-coupling theory} (Oxford University Press, Oxford, 2008).

\bibitem{gleim_00}
T. Gleim and W. Kob, 
Eur. Phys. J. B {\bf 13}, 83 (2000).

\bibitem{Han06} J.-P. Hansen and I. R. McDonald, \emph{Theory of Simple Liquids} (Elsevier Science, 2006).

\bibitem{chaudhuri_07}
P. Chaudhuri, L. Berthier, and W. Kob,
Phys. Rev. Lett. {\bf 99}, 060604 (2007).

\bibitem{Pli95} S. Plimpton, J. Comput. Phys. {\bf 117}, 1 (1995)



\end{thebibliography}
\end{document}